\documentclass[review]{elsarticle}

\usepackage{hyperref}
\usepackage{graphicx}
\usepackage{subcaption}
\usepackage{amsmath}
\usepackage{multirow}
\usepackage{changepage}
\usepackage[normalem]{ulem}

\journal{arXiv}

\begin{document}

\begin{frontmatter}

\title{Visual Analyses of Music History:\\A User-Centric Approach}

\author{Jingxian Zhang}
\ead{jingxian@mit.edu}
\address{Media Lab, Massachusetts Institute of Technology, Cambridge, MA 02139, USA}

\author{Dong Liu\corref{mycorrespondingauthor}}
\cortext[mycorrespondingauthor]{Corresponding author}
\ead{dongeliu@ustc.edu.cn}
\address{CAS Key Laboratory of Technology in Geo-Spatial Information Processing and Application System, University of Science and Technology of China, Hefei 230027, China}

\begin{abstract}
Music history, referring to the records of users' listening or downloading history in online music services, is the primary source for music service providers to analyze users' preferences on music and thus to provide personalized recommendations to users. In order to engage users into the service and to improve user experience, it would be beneficial to provide visual analyses of one user's music history as well as visualized recommendations to that user. In this paper, we take a user-centric approach to the design of such visual analyses. We start by investigating user needs on such visual analyses and recommendations, then propose several different visualization schemes, and perform a pilot study to collect user feedback on the designed schemes. We further conduct user studies to verify the utility of the proposed schemes, and the results not only demonstrate the effectiveness of our proposed visualization, but also provide important insights to guide the visualization design in the future.
\end{abstract}


\end{frontmatter}

\section{Introduction}
%music download history visualization is important
Online content distribution services have been widely accessed by Internet users for viewing and downloading digital content. To enhance user satisfaction and loyalty, service providers often analyze the users' activity history to mine the interests and preferences of users, and utilize the recommender systems to provide personalized recommendations of content that suits a user's interests \cite{adomavicius2005toward}. In this paper, we focus on the online music services as an exemplar of content distribution services; accordingly, we refer \emph{music history} to the records of users' listening or downloading history, the primary source for analyses and making recommendations.

% Downloading digital content from online content distribution services has been a common practice for Internet users. To enhance user satisfaction and loyalty, service providers often analyze the users' download history to mine the interests and needs of users, and utilize the recommender systems to provide personalized recommendations for content that suit a user's interest \cite{adomavicius2005toward}. In online music services, music history, which refers to the records of users' listening or downloading history in the platform, is the primary source for such analyses.

%how current music providers do recommendation
Providing personalized recommendations has been a key feature in many existing online music services to improve user experience. The currently adopted recommender systems can be divided into two dimensions: content-based filtering that recommends content items similar to what the user has liked before, and collaborative filtering (CF) that utilizes not only the past behavior of one user but also of other users \cite{goldberg1992using}. Practical systems often combine the content-based filtering and CF approaches to achieve better performance \cite{adomavicius2005toward}. Besides, recent studies reveal that user interests are not static but rather dynamic, and thus user interest drifts have been taken into account in designing adaptive recommendation systems \cite{abel2011analyzing,cao2009enhancing,koren2010collaborative}.

%user interactive recommendations, visual analyses - knowing user needs and perception
Although automated recommender systems have achieved remarkable success, it has been noticed that enabling user interactive recommendations is still quite important \cite{bogdanov2013semantic,bostandjiev2012tasteweights,parra2014see}. On one hand, the recommendation algorithms may be fine tuned by using user interactions as feedback; on the other, users feel the interactive recommendations more understandable, controllable, and personalized. Obviously, visualization plays an essential role in the interactive recommendations. Since almost all the recommender systems depend on users' historical data, it is naturally to ask whether users' data can be visualized, and then recommendations based on the visualization can be more understandable and more acceptable to the common users. In addition, visual analyses of user behavior also provide great help to service providers, as visualization capably supports the open-ended exploration and provides assistance to analysts \cite{basole2013understanding}.

However, the visual analyses of users' historical data in online services were not well studied yet. There exist several challenges in visualizing such data, including how to display the relevances between the content items that one user had accessed, how to display the history as temporal data with severe temporal discontinuity, how to evaluate the quality of visualization both subjectively and objectively, etc. More importantly, such visual analyses are intended for common users, which raises quite a challenge in making them comprehensive and intuitive.

%some challenges in visualizing user behavior
% There are some more challenges in visualizing user behavior for analyses as well as for making recommendations. In view of the possible user interest drifts, it is crucial to display the relevances between the content items that one user had accessed, but how to quantify and visualize the relevances is a difficulty. Displaying temporal and drifting data remains a challenge in visualization, especially taken into account the interpretability and perceivability. Moreover, how to evaluate the visualization design, both objectively and subjectively, is not well studied before.

%what we do in this work
In this paper, we take a user-centric approach to the design of visual analyses of users' music history. We investigate user needs and users' perception on such visual analyses through a pilot study. Then we design four kinds of visualization schemes, and conduct user studies to comparatively evaluate the proposed schemes. Results of the conducted studies not only demonstrate the effectiveness of our proposed visualization, but also provide insights to guide the visualization design in the future.

This paper is an extended version of \cite{Liu2016MMM}, which reports results of the user studies for our designed visualization schemes. Our previous paper \cite{zhang2014visualization} concentrates on the visualization design issues. Compared to \cite{Liu2016MMM,zhang2014visualization}, this paper makes twofold new contributions. First, a new visualization scheme termed Calendar plot is proposed and evaluated in this paper. Second, a pilot study as well as its results are reported in this paper.

%paper structure
The remainder of this paper is organized as follows. Section \ref{sec_relwork} reviews previous work related to this paper. Section \ref{sec_pilot} presents the details of the pilot study. Our proposed visualization schemes are introduced in Section \ref{sec_design}. User studies are reported in detail in Section \ref{sec_us}. Concluding remarks and future work are presented in Section \ref{sec_conc}.

\section{Related Work}\label{sec_relwork}
Researchers have shown increasing attention to visualized and interactive recommendations that help users discover interesting items with understanding the reason of recommendations. Visualization of music history is of great interest recently, leading to some practices of such visualization for online music services. User-centric approach is well recognized in designing visualizations, and how to evaluate visualizations from users' perspective has been a topic studied mostly in an empirical manner. We will briefly review some related work on the above aspects.

\subsection{Visualized and interactive recommendations}
As mentioned, recommender systems can be roughly classified into three categories: content-based filtering \cite{pazzani1997learning}, CF \cite{billsus1998learning}, and hybrid approaches \cite{balabanovic1997fab}. Visual recommendations also fall into one of the three categories. For example, Bogdanov \emph{et al.} \cite{bogdanov2013semantic} proposed a content-based recommendation that infers high-level semantic descriptors from the music tracks of one user, and then utilizes the semantic descriptions to perform recommendations or visualization of user's preferences. PeerChooser \cite{o2008peerchooser} is a collaborative recommender system with an interactive graphical explanation interface, which enables user to select ``similar users" in her own mind. Hybrid visual recommendations are more attended, for example the recommendations from multiple social and semantic web resources \cite{bostandjiev2012tasteweights}, and the visual user-controllable interface that encourages user to manually control the recommendation strategies \cite{parra2014see}. Regarding the type of content, the work in \cite{baur2009pulling} may be the most similar one to our work, as it proposed several visualizations of music history also for recommendations. However, all the above-mentioned work did not consider the underlying user interest drifts in the users' behavior data, therefore failed to display the temporal dynamics of such data. In this paper, we put emphases on the temporal dynamics as inspired by the progresses in making temporally adaptive recommendations \cite{abel2011analyzing,cao2009enhancing,koren2010collaborative}.

\subsection{Visualization of music history}
Visualization of music-related data has been studied in a wide range. Earlier work was done to visualize music collections such as personal music libraries \cite{knees2006innovative,torrens2004visualizing}. The first visualization of music history seems to be presented in \cite{baur2009pulling}, which added the temporal dimension into consideration. Another work, based on the user's music library, generated a humanoid cartoon-like character called Musical Avatar to visualize the user's interests \cite{bogdanov2013semantic}.

There have been some practices of visualization of music history for online music services, e.g. the well-known Last.fm\footnote{http://www.last.fm/}. For example, \emph{Last Chart!}\footnote{http://www.lastchart.com/}, a webapp, provides various kinds of visualization including bubble chart, force-directed graph, and so on, for Last.fm users. A software named \emph{Last.fm Extra Stats}\footnote{http://www.last.fm/group/Last.fm+Extra+Stats} is also available that visualizes user's music history in stacked graphs. Note that these applications are not provided by the service officially, but by communities using open application programming interfaces (APIs) to access data from the service. It shows the user needs to explore music history in an intuitive manner.

Existing work is mostly concerned with the structured metadata of music, e.g. artist, album, genre, but lacks the mining and utilization of implicit relevances between music items. In terms of recommender systems, the existing work corresponds to content-based filtering but not to CF. In this paper, we try to involve collaborative relevances between music items into the visualization so as to make recommendations from the CF approach.

\subsection{Evaluation of visualization}
Evaluation of information visualizations has always been an indispensable part of related researches. Carpendale discussed different types of evaluations as well as their pros and cons \cite{carpendale2008evaluating}. Lam \emph{et al.} \cite{lam2012empirical} focused on empirical studies in information visualization and summarized seven scenarios to discuss what might be the most effective evaluation of a given information visualization. Basole \emph{et al.} \cite{basole2013understanding} proposed a three-phase user study including tutorial, practice, and evaluation for assessing their visualization design. In this paper, we design user studies as a learning-practice-test workflow, similar to that in \cite{basole2013understanding}. However, our purpose of visualization is to enable common users to perform some visual analyses, which seems more difficult than previous studies. Moreover, we perform a pilot study in conjunction with the visualization design, forming a closed loop of user-centric approach.

% \section{Visualizations}

\section{Pilot Study}\label{sec_pilot}
Since the visual analyses of users' music history were not well studied yet, it is very important to understand user needs before the design, for example, which information is required for such visual analyses? The information should be displayed in which manner for better usability? To collect users' opinions, we first prepare some rough ideas as well as conceptual design, and conduct a pilot study where we use the rough ideas to inspire the users, and their feedback is absorbed in polishing the design. This section presents the details of our performed pilot study.

\begin{figure}
\centering
    \begin{minipage}[b]{0.26\textwidth}
    \centering
    \includegraphics[width=0.90\columnwidth]{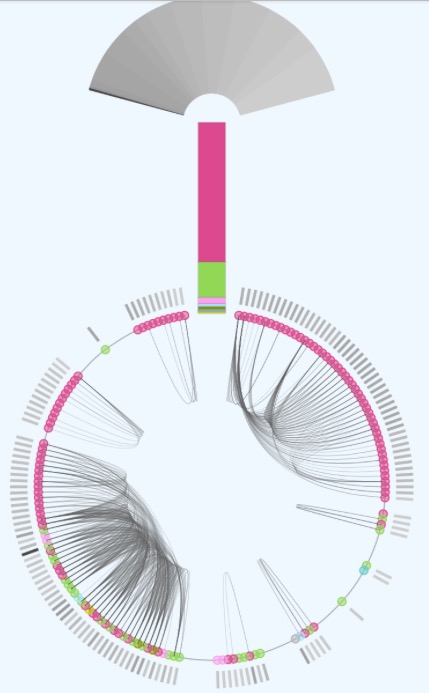}
    % \exedout % first figure itself
    \subcaption{Instrument plot}~\label{fig:pilot_instru}
    \end{minipage}\hfill
    \begin{minipage}[b]{0.30\textwidth}
    \centering
    \includegraphics[width=0.87\columnwidth]{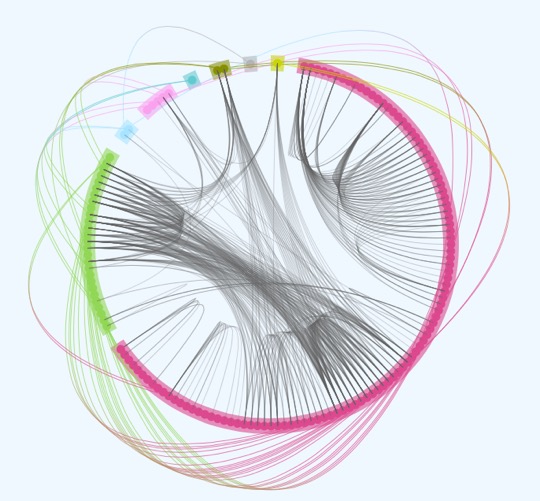}
    % \exedout % second figure itself
    \subcaption{Transitional Pie plot}~\label{fig:pilot_tran}
    \end{minipage}
    \begin{minipage}[b]{0.43\textwidth}
    \centering
    \includegraphics[width=0.97\columnwidth]{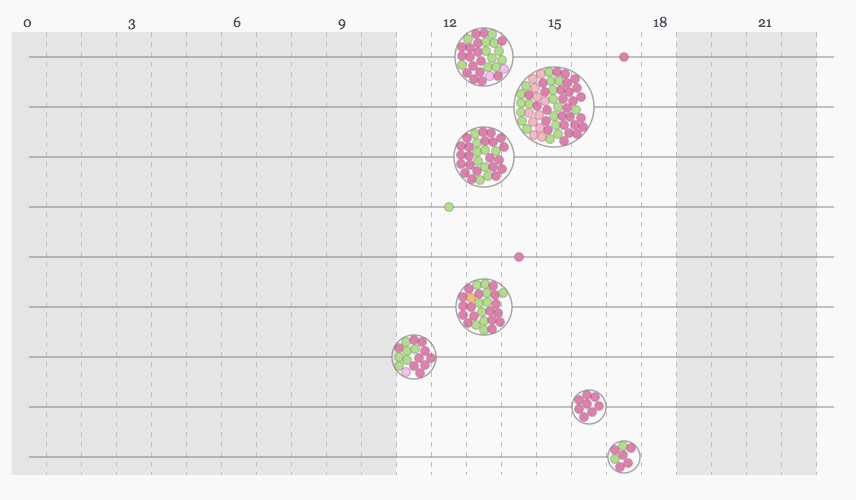}
    % \exedout % second figure itself
    \subcaption{Calendar plot}~\label{fig:pilot_bean}
    \end{minipage}
\caption{Three visualization plots used for pilot study, details in Section \ref{sec_design}.}
~\label{fig:pilot_vis}
\end{figure}

\subsection{Implementation}
10 graduate students (4 females and 6 males) took part in the pilot study. Three questions are prepared for each of them, as shown in Table~\ref{tab:pilot}. After answering Q1, each participant was shown the rough visualization design (as in Figure~\ref{fig:pilot_vis}) together with its description, and then continued answering Q2 and Q3. The details of our visualization design will be given in Section \ref{sec_design}.

\begin{table}[t]
\fontsize{8pt}{10pt}\selectfont
\centering
\caption{Questions in the pilot study}
\label{tab:pilot}
    \begin{tabular}{lp{300pt}}
    \hline\hline
                % & \multicolumn{1}{c}{\textbf{Questions}}\\\hline
    \multirow{3}{*}{Q1} & Which music service/app do you often use? Have you ever seen any kind of visualization of your music history in any  music service/app? If yes, please describe them.\\
    \hline
    \multirow{2}{*}{Q2} & After hearing the description of the three visualization plots, please rank them in order by (1) difficulty of understanding, and (2) usefulness for analyses.\\
    \hline
    \multirow{2}{*}{Q3} & Which features of your music history may you like or dislike to observe if visualized, not limited to those in the three plots?
    \\ \hline\hline
    \end{tabular}
\end{table}

\subsection{Results}
For Q1 in Table~\ref{tab:pilot}, participants mentioned Spotify\footnote{\url{https://www.spotify.com/us/}}, Pandora\footnote{\url{http://www.pandora.com}}, Kuwo\footnote{\url{http://www.kuwo.cn}}, and iTunes\footnote{\url{http://www.apple.com/itunes/}} as their usual music services/apps. 9 out of 10 participants had never saw visualization of their own music history, only one participant had seen simple histograms of listening times of music items. Overall, there is a lack of visual analyses tools for music history that are generally accepted by common users.

For Q2, 9 out of 10 participants thought Figure~\ref{fig:pilot_vis} (c) was the easiest one to understand. Within these 9 participants, 7 of them prefer Figure~\ref{fig:pilot_vis} (a) than Figure~\ref{fig:pilot_vis} (b) as easier, and the rest 2 prefer (b) than (a); moreover, 6 of them chose Figure~\ref{fig:pilot_vis} (c) to be the most useful one, and the rest 3 chose Figure~\ref{fig:pilot_vis} (a). One participant chose Figure~\ref{fig:pilot_vis} (b) to be both the easiest one and the most useful one. There seems a consistency between easiness and utility in users' mind, i.e. user feels like a visualization is useful because it is easy to understand.

For Q3, the results are summarized in Table~\ref{tab:pilot_q3}. Note that the three visualization plots in Figure~\ref{fig:pilot_vis} cover the following features of user's music history: genre (in (a), (b), and (c)), release year of music track (in (a)), relevances between tracks (in (a) and (b)), and daily pattern of music (in (c)). Regarding relevances between music tracks, 4 out of 10 participants could not understand them easily and thus preferred not to observe them; however, one participant was very interested in them and would like to observe more details of how these relevances are calculated and how they can be used for recommendations. Besides, participants of the pilot study also mentioned several features that they may like to observe, especially the recommendations based on music history. Participants reported willingness to get different recommendations based on different criteria, for example, based on the overall music history, or a specific portion of history. Also, social features seem to be well welcomed, e.g. comparison with friends or find similar users.

\begin{table}
\fontsize{8pt}{10pt}\selectfont
\centering
\caption{Features that users may or may not like to observe in the visualizations}
\label{tab:pilot_q3}
\begin{tabular}{lcc}
\hline
\textbf{Features} & \textbf{\#May like} & \textbf{\#May dislike} \\\hline
\textbf{a.} genre & 5 & 0 \\
\textbf{b.} daily pattern of music listening & 4 & 1 \\
\textbf{c.} recommendations based on history & 3 & 0 \\
\textbf{d.} favorite tracks & 3 & 0 \\
\textbf{e.} comparison with my friends & 2 & 0 \\
\textbf{f.} find similar users & 2 & 0 \\
\textbf{g.} albums listened recently & 2 & 0 \\
\textbf{h.} artists' photos & 1 & 0 \\
\textbf{i.} release year  & 1 & 1 \\
\textbf{j.} relevances between tracks & 1 & 4 \\
\textbf{k.} location of music listening & 1 & 0 \\  \hline
\end{tabular}
\end{table}

\subsection{Insights}
Based on the results of pilot study, we revise the visualization design as will be detailed in Section \ref{sec_design}. Here we would like to provide some insights over user needs in such visualizations.

We find that users are interested in seeing their music history visualized. Visual recommendation based on music history is highly praised by users. Importantly, users themselves understand that recommendations can be made on different pieces of history, revealing that user interest drifts are indeed common. Therefore, how to enable such adaptive recommendations in a visualized manner is worthy of further study.

Users expect such visualizations to be easy to understand and easy to use. Relevances between music tracks, seemingly very important in making recommendations, is not a easy concept to common users. It may need further efforts on presenting such relevances in a more intuitive manner.

Social features have been observed to attract common users, like comparing with friends or finding similar users. Though this paper is focused on visualization of one user's music history, we leave the visualized analyses of multiple users for future work.

\section{Visualization Design}\label{sec_design}
In this section we describe four kinds of visualization plots for music history, namely Bean plot, Transitional Pie plot, Instrument plot, and Calendar plot. The former three plots are used for visual analyses, whereas the Calendar plot can be used for analyses as well as recommendations. We will first discuss the data to be visualized, then introduce the relevances between music tracks, and present the four plots one by one.

\subsection{The Data to Visualize}
The raw data recording music history of one user can be described as pairs of downloading timestamp and the identity of music track. These music tracks along the temporal dimension, as well as the underlying relevances among them, constitute the data to be visualized in our work.
\subsubsection{Sessions}
\begin{figure}[t]
\setcounter{figure}{1}
\centering
\includegraphics[width=.8\textwidth]{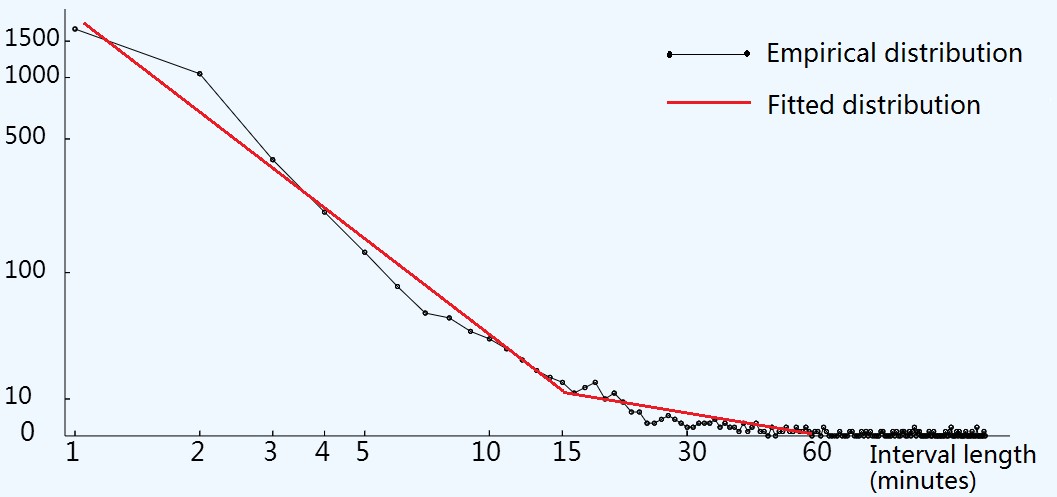}
\caption{Statistics of the length of intervals between successive actions in our data set. The empirical distribution (shown in black) can be well fitted by a piecewise power-law distribution (shown in red).}
\label{fig:timeplot}
\end{figure}
The music history of one user may last for a long period but access actions are distributed discontinuously in the entire history. Specifically, similar to many other online services, we observed \emph{sessions} in the music history of most users, where each session consists of a series of consecutive actions with short intervals in between them, and the intervals between sessions are usually much longer. In order to divide the music history into sessions, we first made some statistics about the intervals between successive actions according to our data set, which are shown in Fig.~\ref{fig:timeplot}. It can be observed that the length of intervals follows a piecewise power-law distribution, and more than 98\% intervals are less than 1 hour. Therefore, we define sessions as if two successive actions have an interval less than 1 hour, then both are in the same session.
\subsubsection{Relevances among music tracks}
In order to depict user interests towards music and especially the drifts of user interests, it is crucial to visualize not only the music tracks but also the relevances among them. We focused on the relevances so as to identify the user interest drifts and to adapt recommendations to the drifts. According to the practice of recommender systems, relevances among the content items may be approached in two ways. The first is inspired by content-based filtering where content similarity is defined on the features of content items, leading to feature-based relevances that are dependent on the adopted features of content. The second is inspired by CF that makes use of the behaviors of the crowd, in this way we defined collaborative relevances.

For music specifically, we have considered \emph{genre} and \emph{release year} as the features of music content in the visualization design. Genre is often taken as a basic categorical feature for music content, and color coding is usually adopted to display genres. Synesthesia may interpret the relations between colors and music genres, as analyzed in some previous work such as \cite{IEEEhowto:Holm09}, which revealed that a default color-genre mapping can be suitable for a given country or region. Release year represents the numerical features of content. We used graylevels to display the release years of music tracks in our visualization plots. These colors and graylevels can impress the intrinsic features of music tracks upon the viewers and thus imply the relevances among music tracks. Note that other features of music content including categorical and numerical features can be visualized in the same manners, but how to display the acoustic features of music is an open problem, and it raises several challenges to visualize multiple features simultaneously \cite{IEEEhowto:kehrer13}. We leave the other features to our future work.

Instead of relying on content analyses to recommend similar items, CF utilizes the ``crowd wisdom'' to provide recommendations. The basic assumption of CF is that if many people like both items $a$ and $b$, then a new user in favor of $a$ probably likes $b$ as well. In another view, we may argue that the more users like both $a$ and $b$, the higher relevance or similarity is between $a$ and $b$. We name such relevances as collaborative relevances.

Collaborative relevances (CR) have some advantages over the abovementioned feature-based relevances. On the one hand, CR are not dependent on the content itself, thus CR are generally available for different kinds of content without the requirement of content analyses that may be difficult or time-consuming for multimedia content. On the other, CR help to discover unexpected and interesting patterns of user behaviors, which may be difficult to interpret or predict from the content features. Therefore, CF has achieved great successes in making recommendations and CR were studied for visualization in previous work e.g. \cite{baur2009pulling}.

In this paper, we have made two remedies on the basic definition of CR. Firstly, the relevance between items $a$ and $b$ is dependent on the number of users who accessed both items \emph{within a short period}, this time constraint is due to the consideration of possible user interest drifts, as we may reasonably assume that user interest does not change within a short period. Secondly, the time constraint will make the relevances quite sparse among many items, thus we augment them with indirect collaborative relevances to avoid data scarcity.

Our definitions of CR are as follows. Let $N$ be the number of users and $S_i,i=1,\cdots,N$ be the set of tracks that user $i$ had accessed. For the track $a\in S_i$, $t_i(a)$ denotes the timestamp of the access action. Then for two tracks $a$ and $b$, we define an indicator function as
\begin{eqnarray}
r_i(a,b)=
\begin{cases}
1 & \mbox{if } a\in S_i,b\in S_i,\\
  & |t_i(a)-t_i(b)|\leq T_0\\
0 & \mbox{otherwise}
\end{cases}
\end{eqnarray}
where $T_0$ equals 1 hour according to the analyses presented before. Moreover, we define direct collaborative relevance between $a$ and $b$ as
\begin{equation}
R_D(a,b)=\sum_{i=1}^N r_i(a,b)
\end{equation}
and indirect collaborative relevance as
\begin{equation}
R_I(a,b)=\sum_{i=1}^N \sum_{x\neq a,b} (r_i(a,x)+r_i(b,x))
\end{equation}
And finally we define the collaborative relevance as
\begin{equation}
R(a,b)=R_D(a,b)+\lambda R_I(a,b)
\end{equation}
where $\lambda$ is a constant and was set to 0.25 empirically.

\subsection{Bean plot, Transitional Pie plot, and Instrument plot}
\begin{figure}[!t]
\centering
    \begin{minipage}[b]{0.49\textwidth}
    \centering
    \includegraphics[width=0.99\columnwidth]{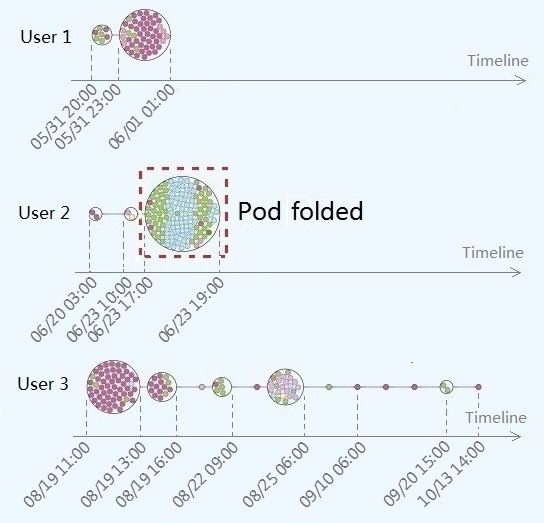}
    % \exedout % first figure itself
    \subcaption{A folded Bean plot}
    ~\label{fig:vis_bean1}
    \end{minipage}\hfill
    \begin{minipage}[b]{0.49\textwidth}
    \centering
    \includegraphics[width=0.99\columnwidth]{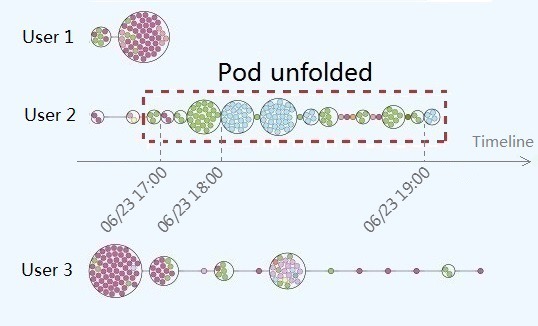}
    % \exedout % second figure itself
    \subcaption{An unfolded Bean plot}
    ~\label{fig:vis_bean2}
    \end{minipage}
\caption{(Left) Bean plot showing the music history of three users. Each small, color-filled circle (named a bean) represents a music track and each larger disc (named a pod) represents a download session. Colors of beans stand for genres. Pods are arranged in chronological order. (Right) Bean plot provides interactive display that one pod being clicked will unfold to multiple smaller pods to represent subsessions, each of which has a single genre.}~\label{fig:vis_bean}
\end{figure}

\begin{figure}[!t]
\centering
    \begin{minipage}[b]{0.9\textwidth}
    \centering
    \includegraphics[width=0.99\columnwidth]{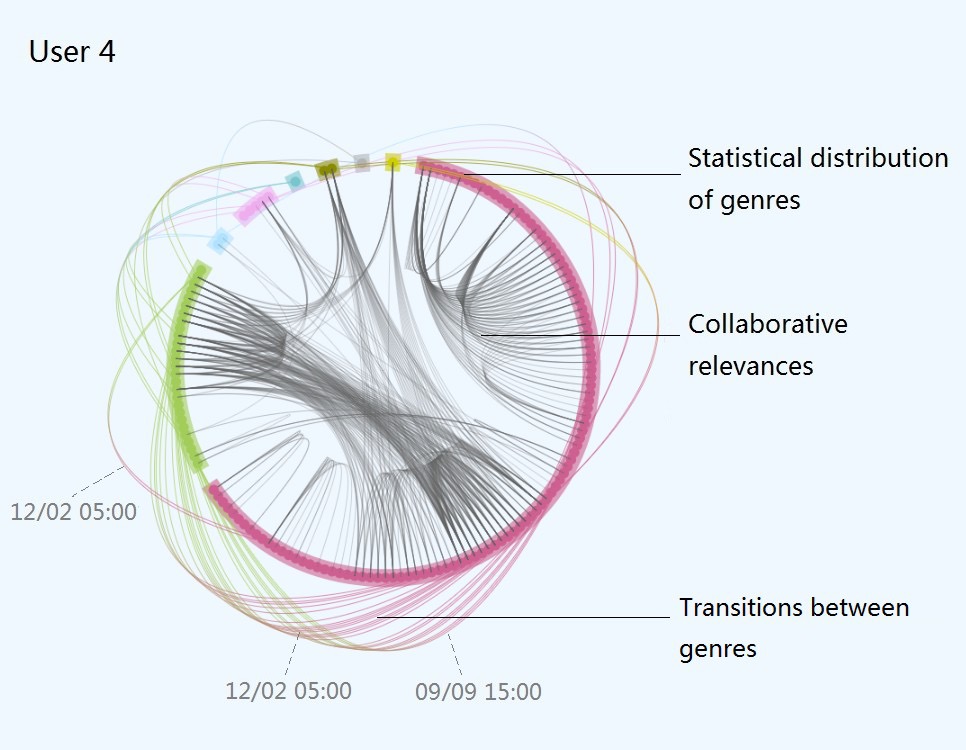}
    \end{minipage}
\caption{Transitional Pie plot showing the music history of one user. Similar to pie chart, the disc shows the proportions of different genres. Within each genre, tracks are arranged in chronological order so that each downloaded track has a corresponding position on the disc. Bezier curves inside the disc display the collaborative relevances among music tracks. Bezier curves outside the disc show the transitions between genres, two successively downloaded tracks of different genres are connected by a outer-disc curve.}
~\label{fig:vis_tran}
\end{figure}

\begin{figure}[!t]
\centering
    \begin{minipage}[b]{0.51\textwidth}
    \centering
    \includegraphics[width=0.99\columnwidth]{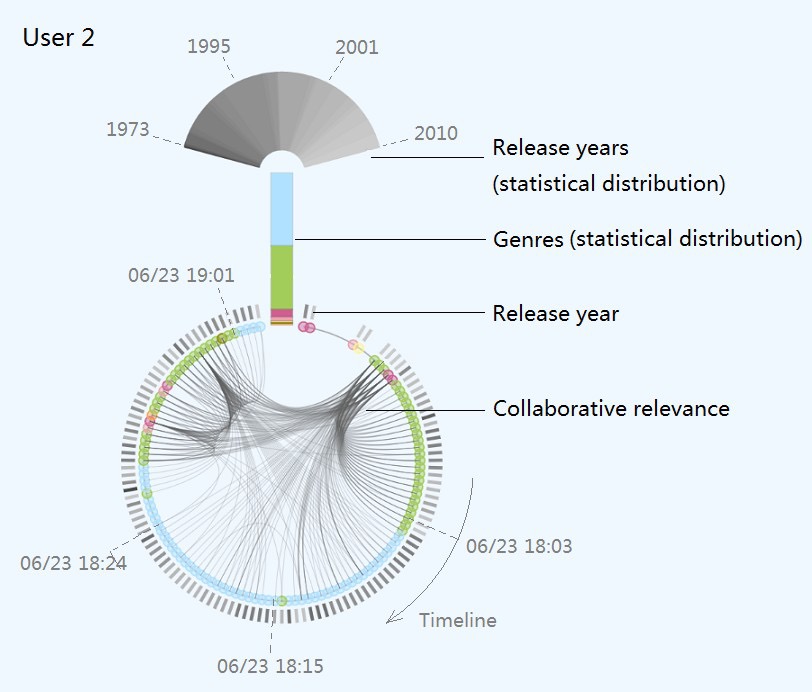}
    % \exedout % first figure itself
    % \subcaption{}
    ~\label{fig:vis_instru1}
    \end{minipage}\hfill
    \begin{minipage}[b]{0.49\textwidth}
    \centering
    \includegraphics[width=0.99\columnwidth]{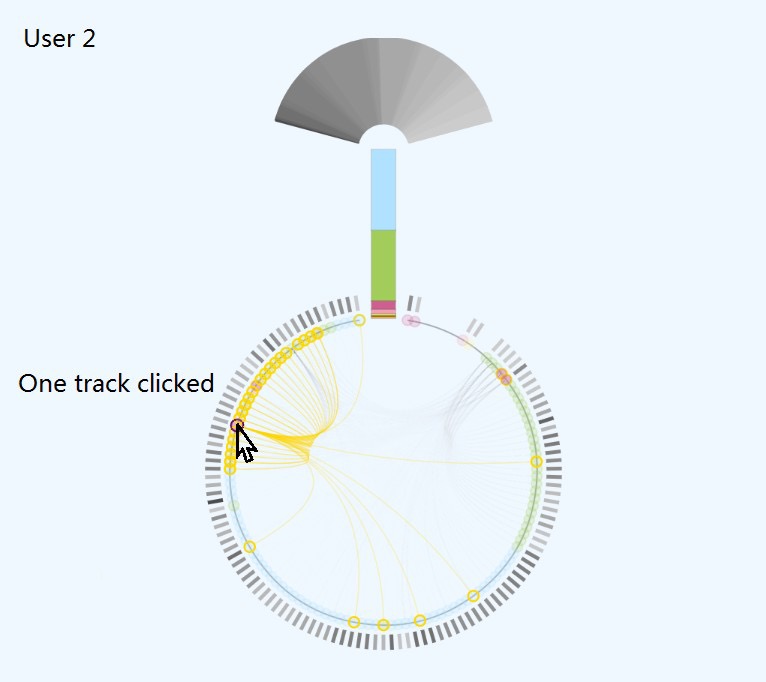}
    % \exedout % second figure itself
    % \subcaption{}
    ~\label{fig:vis_instru2}
    \end{minipage}
\caption{(Left) Instrument plot showing the music history of one user. The timeline is represented by a disc as the body of the instrument, where the music tracks are arranged in chronological order. The gray bars alongside the tracks indicate release years. Bezier curves inside the disc represent the collaborative relevance among music tracks. The distributions of release year and genre are shown as the headstock and the neck of the instrument, respectively. (Right) Instrument plot provides interactive display, once a track is clicked, all its related tracks will be highlighted.}~\label{fig:vis_instru}
\end{figure}

For visual analyses of music history especially the relevances between music tracks, we had proposed Bean plot, Transitional Pie plot, and Instrument plot in our previous work \cite{zhang2014visualization}. Here we enclose them in Figures \ref{fig:vis_bean}, \ref{fig:vis_tran}, and \ref{fig:vis_instru} for completeness of this paper. Please refer to the captions of these figures and also \cite{zhang2014visualization} for more details.

\subsection{Calendar plot}
\begin{figure}[!t]
\centering
    \begin{minipage}[b]{0.99\textwidth}
    \centering
    \includegraphics[width=0.99\columnwidth]{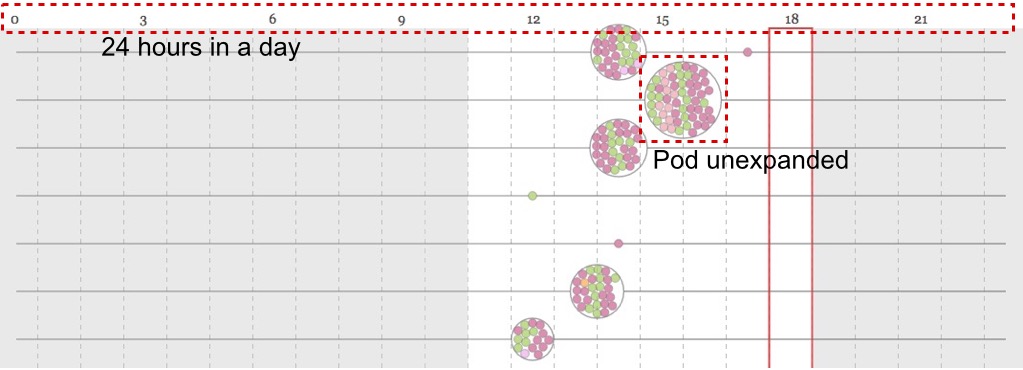}
    % \exedout % first figure itself
    % \subcaption{}
    ~\label{fig:vis_modibean1}
    \end{minipage}\hfill
    \begin{minipage}[b]{0.99\textwidth}
    \centering
    \includegraphics[width=0.99\columnwidth]{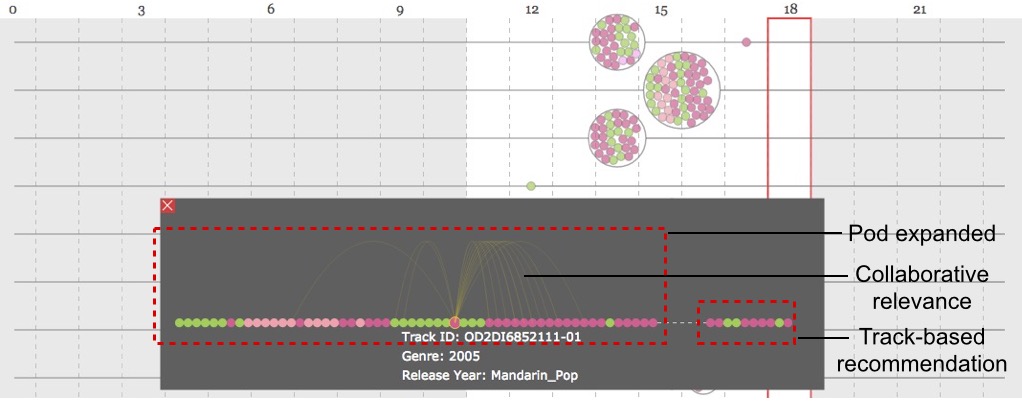}
    % \exedout % second figure itself
    % \subcaption{}
    ~\label{fig:vis_modibean2}
    \end{minipage}
\caption{(Top) Calendar plot showing the music history of one user. Horizontal axis shows the 24 hours of a day and each row stands for one day. Tracks in a session are placed into a pod, similar to the Bean plot. (Bottom) When a pod is clicked, beans in this pod are expanded onto a line in a small window. Details of each music track and the collaborative relevances will be displayed with mouse hovering. When a track is clicked, recommendations based on that track will be displayed at the end of the line.}~\label{fig:vis_modibean}
\end{figure}

In this paper, we further propose a new visualization plot named Calendar plot for visual analyses and recommendations. Calendar plot, shown in Figure~\ref{fig:vis_modibean}, adopts two-layer presentation: the first layer (unexpanded) is similar to Bean plot, except that pods are now organized into a calendar view to emphasize daily patterns as well as temporal adaptive recommendations; the second layer (expanded) displays more details within a session, the information of each track such as track ID, genre, release year, is shown upon mouse hovering the corresponding bean. The collaborative relevances among music tracks are also highlighted. Moreover, it provided recommendations based on a single track once the track is clicked.

\begin{figure}[!t]
\centering
    \begin{minipage}[b]{0.99\textwidth}
    \centering
    \includegraphics[width=0.99\columnwidth]{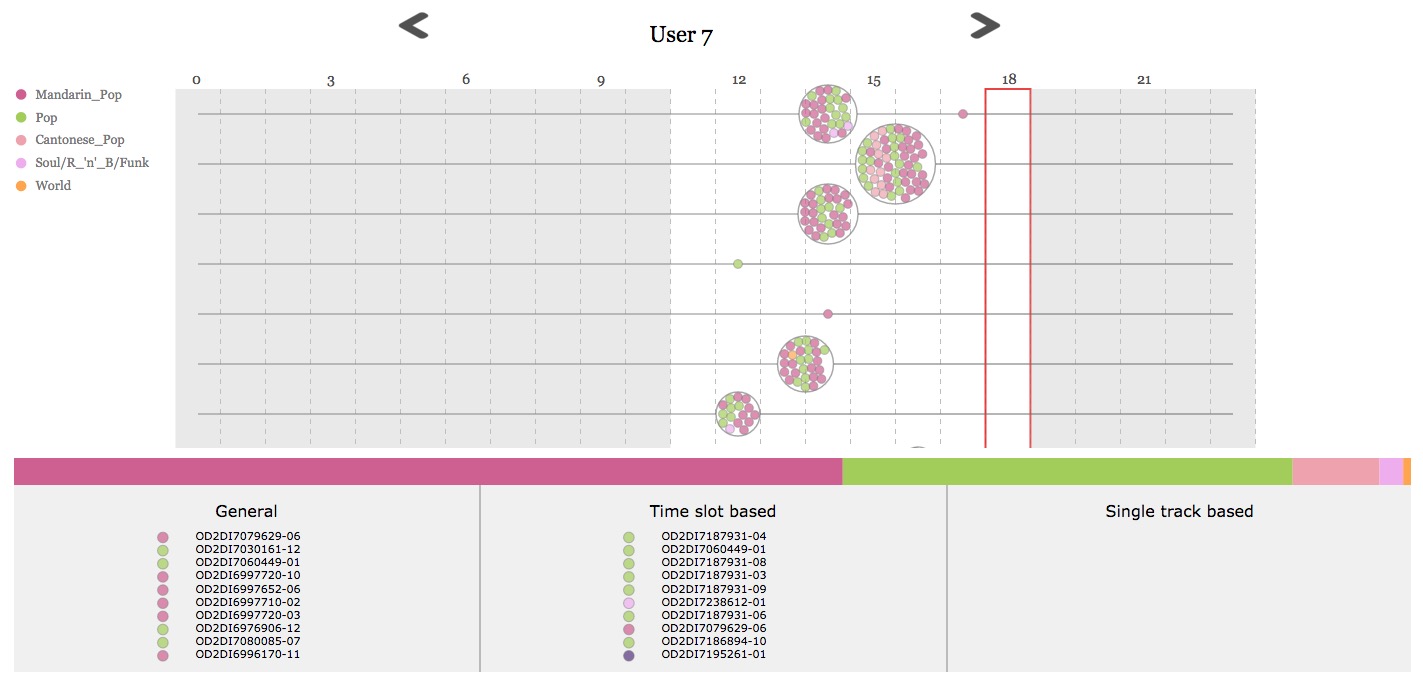}
    \end{minipage}
\caption{Different recommendations summarized in the Calendar plot, including recommendations based on the entire history (General), based on the given period (Time slot based), and based on a single track (Single track based, will be displayed when a track is clicked). Due to privacy concern, the detailed information of music is replaced with IDs in this figure.}
~\label{fig:vis_modibean_interface}
\end{figure}

Figure~\ref{fig:vis_modibean_interface} displays different kinds of recommendations that are all visualized in our proposed Calendar plot. First, general recommendations are provided based on the entire music history of one user. Second, once a time slot is selected--currently each hour within a day is a basic time slot--recommendations are provided based on the history within that time slot; this design comes from the pilot study showing that users are interested in their daily patterns. Third, recommendations based on a single track can be made visible as long as that track is selected, and such recommendations are also provided visually in Figure~\ref{fig:vis_modibean}.

\section{User Studies}\label{sec_us}
We conduct user studies to evaluate the usability of three of our proposed visualization plots: Bean plot, Instrument plot, and Transitional Pie plot. Note that these three plots are used for visual analyses and thus directly comparable. The Calendar plot, on the contrary, further provides visual recommendations. How to evaluate the user experience toward recommendations is still an open problem \cite{adomavicius2005toward}, therefore, we first study the usability of visual analyses.

In the studies, participants complete a learning-practice-test workflow. Participants are required to first learn the design of the three plots; then after practice, they are asked to analyze the plots of new users and to answer both specific questions and open-ended questions regarding the user interest drifts; finally, questionnaire and survey are performed to collect the participants' feedback about the experience of using the visualization plots.

\subsection{Implementation}
We have implemented and tested the proposed visualization design with a real-world data set provided by an online music service. Data preprocessing is performed offline, including the division of sessions and the calculation of collaborative relevances. The visualization plots of each user are drawn upon request in a webpage view, implemented by HTML5 and JavaScript, which also enables the designed user interactions. Online rendering of the plots is computationally efficient and does not incur noticeable delay in mainstream web browsers.

15 undergraduate students (6 females and 9 males) participate the user studies. Participants have different majors including science and engineering, with ages ranging from 20 to 23 (mean: 21.6). Participants reported different levels of interests in online music services, but none of them had experience of visual analyses.

\subsection{Tasks}
Each participant undertakes the user study in 4 sessions: learning, practice, test, and questionnaire. The learning session is to help participants understand the proposed visualization design. In the learning session, the instructor briefly introduces background and the data, explains the three plots in detail, and then shows a screen-captured video to display the user interface as well as interactions. Questions from participant are encouraged at any time and will be answered immediately. The learning session lasts for around 15 minutes for one participant.

\begin{table*}[t]
\fontsize{8pt}{10pt}\selectfont
\begin{center}
\caption{Questions in the Practice Session}
\label{tab:practice}
\begin{tabular}{lp{200pt}}
  \hline\hline
  \multirow{6}{*}{Bean plot} & PQ1: How many subsessions are there in the session A of user 2? \\

  & PQ2: Whose download history is more continuous in time, user 2 or user 3? \\

  & PQ3: Whose download history is more consistent in music genre, user 7 or user 8?\\
  \hline
  \multirow{7}{*}{Transitional Pie plot} & PQ4: Whose download history has more genres, user 12 or user 18? \\

  & PQ5: Which genre is the most lasting in user 4's download history? \\

  & PQ6: Combining the collaborative relevances and the transitions, whose interest lasts for longer time, user 2 or user 3?\\
  \hline
  \multirow{6}{*}{Instrument plot} & PQ7: Who prefers music tracks with older release year, user 2 or user 5? \\

  & PQ8: List the music tracks that have the highest relevances in user 2's download history.\\

  & PQ9: Find the track that has relevances with the most kinds of genres in user 27's download history.\\

  \hline\hline
\end{tabular}
\end{center}
\vspace{-5mm}
\end{table*}

The practice session is then conducted to enhance the comprehension and familiarity of the participant on the plots. In this session, participant is asked to analyze some provided plots and answer some questions (see Table~\ref{tab:practice} for the questions) based on the plots. For example, the instructor provides the Bean plots of user 7 and user 8, and then asks the participant: ''whose music history is more consistent in music genre, user 7 or user 8?" (PQ3 in Table~\ref{tab:practice}) Instructor will answer any question of the participant, and will explain the plots again to the participant, if necessary. The practice session lasts for around 19 minutes for one participant.

\begin{table*}[t]
\fontsize{8pt}{10pt}\selectfont
\begin{center}
\caption{Questions in the Test Session}
\label{tab:test}
\begin{tabular}{lp{200pt}}
  \hline\hline
  \multirow{8}{*}{Specific questions} & Q1-1: Whose download history is the most consistent in time?\\
  & Q1-2: Whose download history is the most consistent in genre?\\
  & Q2: Whose downloaded tracks have the highest relevances?\\
  & Q3-1: Whose interest is the most consistent?\\
  & Q3-2: Whose interest is the least consistent?\\
  \hline
  \multirow{5}{*}{Open-ended questions} & Q4: Write down your findings of user 12's download history as many as possible.\\
  & Q5: Write down your findings of the differences between the download history of user 2 and user 4 as many as possible.\\
  \hline\hline
\end{tabular}
\end{center}
\vspace{-5mm}
\end{table*}

The test session is the most important part of the user study. In the test session, the visualization plots of 10 users\footnote{http://staff.ustc.edu.cn/\%7Edongeliu/stuff/userStudyPlots.pdf} are provided to participant for visual analyses. Participant is asked to analyze all the plots and then to answer 3 specific questions and 2 open-ended questions (the questions are given in Table~\ref{tab:test}). In the test session, instructor does not offer any assistance to the participant, neither tell the participant which plot to look at, nor answer any question regarding the plots. The only instruction is to encourage the participant to answer the open-ended questions as comprehensive as possible. By this design, we hope to verify whether the participant has learnt the characteristics of different plots and can choose the plots for visual analyses, and to observe what the participant can find from the exploration of the plots. The test session lasts for around 37 minutes for one participant.

The last session is to ask the participant to finish a five-point Likert-scale questionnaire regarding the experience of using the plots for visual analyses. Also, a quick survey is conducted to collect the participant's feedback on the visualization design. In total, the user study for one participant lasts for 71 minutes on average. All participants are instructed by the same investigator.

\subsection{Results}
\textbf{Practice Session} In the practice session, we ask the participants to answer some specific questions (given in Table~\ref{tab:practice}) regarding the visual analyses of the provided plots. Participants' answers show a good consistence: for 5 out of 9 questions (i.e. PQ1, PQ4, PQ5, PQ7, PQ9), all the 15 participants give the same answer; for the other 3 questions (PQ2, PQ3, PQ8), 14, 14, and 13 participants give the same answer, respectively. The only exception is PQ6: for this question, 7 out of 15 participants choose user 2 and the other 8 participants choose user 3. These results show that the three plots are not difficult to understand for the participants except for the collaborative relevances and transitions that are shown simultaneously in the Transitional Pie plot, which may lead to diverse understandings.
\\

\textbf{Test Session} In the test session, we ask the participants to answer both specific questions and open-ended questions (given in Table~\ref{tab:test}) after their analyses of the visualization plots of 10 users. These questions are believed to be much more difficult and more subjective compared to the questions in the practice session. On one hand, the analyst shall compare the plots of 10 users to give out the answers to the specific questions. On the other, the open-ended questions indeed have diverse correct answers.

For the specific questions shown in Table~\ref{tab:test}, Q1-1, Q1-2, and Q2 all receive quite consistent answers from the participants. For Q1-1, 13 participants select user 6 and the other two select user 24, which can be verified by looking at the Bean plots of these two users. For Q1-2, 11 participants select user 2, two select user 3, and the other two select user 4, which has been made visible in the Bean plots and Instrument plots. For Q2, observing the Instrument plots or Transitional Pie plots, 10 participants select user 6, and the other five select user 24. Note that such consistence is not trivial, since the participants need to pick one user from the 10 users as the answer to these questions, and the interest patterns of several users are very similar. Therefore, we believe that the plots have depicted the characteristics of user interests that help participants make the correct analyses.

\begin{figure}[!t]
\centering
\includegraphics[width=0.97\columnwidth]{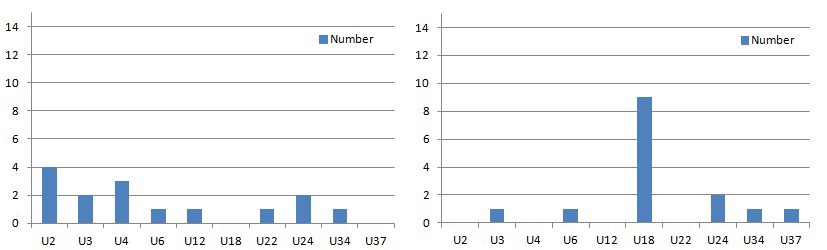}
\caption{The answers to Q3-1 (left) and Q3-2 (right) in the test session. Please refer to the questions shown in Table~\ref{tab:test}} \label{fig:histo}
\end{figure}

The answers to Q3-1 and Q3-2 (summarized in Figure~\ref{fig:histo}) show more diversity among participants. These two questions are not specific to time, genre, or collaborative relevance, but require the participants to present their own understandings of ``user interest drifts,'' which is quite subjective. Participants report difficulty in considering genre and collaborative relevance together when looking for the user with the most or the least consistent interest. This is not surprising since genre and collaborative relevance are displayed separately in our designed plots. This issue shall be addressed in the future work.

\begin{table}
\fontsize{8pt}{10pt}\selectfont
\centering
\caption{The Answers to the Open-Ended Questions in the Test Session}
\label{tab:test2}
\begin{subtable}{.99\linewidth}\centering
\subcaption{Answers to Q4}\label{tab:test_a}
\begin{tabular}{lc} \hline
\multicolumn{1}{c}{\textbf{Findings}} & \begin{tabular}[c]{@{}l@{}} \textbf{\#Mentioned} \end{tabular} \\ \hline
\begin{tabular}[c]{@{}l@{}}The user prefers music tracks with modern release years.\end{tabular} & 14 \\
The user prefers mandarin pop music. & 12 \\
The downloaded tracks have high collaborative relevances. & 10 \\
\begin{tabular}[c]{@{}l@{}}The user's downloading actions are concentrated in time.\end{tabular} & 5 \\
The user's interest is not consistent in genre. & 4 \\
The user likes only a few genres. & 3 \\
There are long intervals between the user's downloading sessions. & 3 \\
The user's interest is quite stable within each session. & 2 \\
\begin{tabular}[c]{@{}l@{}}In different sessions, the numbers of tracks vary quite a lot.\end{tabular} & 2 \\
\begin{tabular}[c]{@{}l@{}}Transitions between genres only happen among tracks\\ that have collaborative relevances.\end{tabular} & 2 \\
\begin{tabular}[c]{@{}l@{}}The release years of the tracks are relatively concentrated.\end{tabular} & 1 \\
The user might have different moods in different sessions. & 1 \\
\textit{The downloaded tracks have low collaborative relevances.} & 1 \\ \hline \end{tabular}
\end{subtable}

\begin{subtable}{.99\linewidth}\centering
\subcaption{Answers to Q5}\label{tab:test_b}
\begin{tabular}{lc} \hline
\multicolumn{1}{c}{\textbf{Findings}} & \begin{tabular}[c]{@{}l@{}} \textbf{\#Mentioned} \end{tabular} \\ \hline
\begin{tabular}[c]{@{}l@{}}User 2 prefers music tracks with older release years.\end{tabular} & 14 \\
\begin{tabular}[c]{@{}l@{}}User 2's interest is more consistent in time and in genre.\end{tabular} & 14 \\
User 2 likes pop and soul music while user 4 likes mandarin pop. & 11 \\
User 4's downloaded tracks have higher collaborative relevances. & 8 \\
User 2 has more consistent interest than user 4. & 5 \\
User 2's downloading sessions contain more tracks. & 3 \\
User 4 likes more genres than user 2. & 2 \\
\begin{tabular}[c]{@{}l@{}}User 4's taste seems to be of popular mass compared to user 2.\end{tabular} & 1 \\
\begin{tabular}[c]{@{}l@{}} \textit{User 2's downloaded tracks have higher collaborative} \textit{relevances.} \end{tabular} & 1 \\ \hline
\end{tabular}
\end{subtable}
\end{table}

The open-ended questions Q4 and Q5 (also shown in Table~\ref{tab:test}) ask the participants to write down as many as possible their findings of one user's music history and of comparison between two users'. Participants' findings are summarized in Table~\ref{tab:test2}. Overall, these findings achieve a good consistence among different participants and at the same time exhibit subjective diversity. There are only two findings inconsistent with the others (typeset italic in Table~\ref{tab:test2}), each approved by only one participant, and both relate to collaborative relevances. It reveals that few participants have misunderstanding on the collaborative relevances, which seem a less straightforward concept for them. Moreover, participants have made findings at different aspects including the release year and genre of music, the download sessions, and the collaborative relevances. Interestingly, participants mention some findings that are beyond our previous analyses. For example, ``transitions between genres only happen among tracks that have collaborative relevances,'' which is mentioned by two participants, was not easy to find at the first sight. Also, some findings are more conjectured than analyzed, e.g. ``the user might have different moods in different sessions, since the relevances between tracks within each session is high, but cross-session relevances are almost none,'' said the participant. Last but not the least, the findings of comparison between two users (answers to Q5) are more consistent among different participants compared to the answers to Q4, which implies that the visual comparison may be easier and more obvious than the visual analysis of single user.
\\

\begin{table}
\centering
\caption{Questions and Answers in the Questionnaire and Survey Session}
\label{tab:survey}
\begin{subtable}{.57\linewidth}
\begin{tabular}[width=0.90\columnwidth]{l} \hline
\textbf{Questions}                                       \\ \hline
The plots are easy to learn. \\
The plots are easy to use. \\ \hline
Relevance is helpful in analyses. \\
The plots make user interests obvious. \\
The designed interactions are intuitive. \\ \hline
 Redundancy exists in Bean plot. \\
 Redundancy exists in Instrument plot. \\
 Redundancy exists in Transitional Pie plot. \\ \hline
 I am willing to use the plots to analyze. \\ \hline
\end{tabular}
\end{subtable}
\begin{subtable}{.42\linewidth}\centering
\begin{adjustwidth}{0.4cm}{}
\centering
\includegraphics[width=1.06\columnwidth]{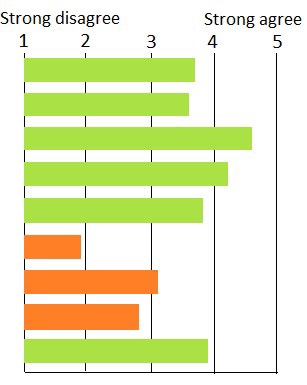}
\end{adjustwidth}
\end{subtable}
\end{table}

\textbf{Questionnaire and Survey Session} After practice and test sessions, we ask the participants to finish a five-point Likert-scale questionnaire that consists of 9 questions shown in Table~\ref{tab:survey}, which also shows the average score of answers to each question. We also conduct a survey to collect participants' feedback.

\textit{Questionnaire.} According to the questionnaire, participants feel the visualization interface is easy to learn and use. This is also verified by the fact that all participants, having no experience of visual analyses, can finish the entire user study with good performance in less than 90 minutes. Regarding the usability of the visualization, participants agree that collaborative relevance is helpful in analyzing user interests, and all of them indeed take the collaborative relevance into consideration during the test session. 13 out of 15 participants agree that the three plots make user interests obvious, and the other two are neutral to this question. Most participants think the designed interactions are intuitive, and we have observed that all participants learn the interactions in Instrument plot very quickly and use the highlighting feature effectively; the interactions in Bean plot are also not a difficulty for the participants, but three of them need some time to understand the concept of subsessions (unfolded pods), once understood, they all use the interactions well in the test session. Furthermore, on whether there is redundancy in the three plots, Bean plot is believed to have none, but Instrument plot and Transitional Pie plot are believed to have some redundancy to some extent by several participants. The reason may be that the Instrument plot displays the genre and release year of each track as well as the statistics of them, and the Transitional Pie plot uses gradient color for the transitions between genres. Finally, participants report willingness to use the visualization plots (9 agree, 5 neutral, and 1 disagree).

\textit{Survey.} In the survey, participants are asked about their opinions on the visualization design. For Bean plot, although it does not display collaborative relevance, its distinctive layout is acknowledged by the participants and it is utilized frequently in the test session. Almost all participants believe that Instrument plot displays the most information and the most important information; participants express an overwhelming preference on the Instrument plot in the test session. The Transitional Pie plot is endorsed by some participants but disliked by some others. Several participants are fond of the Transitional Pie plot as ``the inner- and outer-disc curves can be jointly considered,'' but some others think the Transitional Pie plot can be covered by the Instrument plot and thus is unnecessary. Moreover, participants provide comments and suggestions on the plots. For example, the shape of beans can be changed to square so that the layout of Bean plot may look more regular; the color coding of genres may be adaptive for each user to better distinguish different genres; and so on. These issues will be addressed in our future work.

\subsection{Discussions}
Comparing the three plots we have in the user study, we find that Instrument plot receives the most preference due to its comprehensiveness. Bean plot is approved when the analysts do not concern collaborative relevance and focus on the sessions in music history. Transitional Pie plot is endorsed by several people but less utilized by some others.

Collaborative relevance is believed to be helpful in the analyses, but people have difficulty in combining feature-based and collaborative relevances at the same time since the plots display them separately. Compared to the feedback we get in the pilot study, participants seem to have higher acknowledgement of collaborative relevances after exploring them in the visualization interfaces in the user study. Nonetheless, it needs further study on how to intuitively interpret collaborative relevance and combine it with other features in visualization.

With the help of our proposed visualization design, non-expert people can learn and use the plots for visual analyses of user interests without much difficulty. As the visualization makes user interests obvious, there is a good consistence among the analyses of different people. At the same time, people also make diverse, sometimes subjective findings from the explorative analyses, which implies the visualization may inspire analysts to further investigate the user interests.

% Given the good performance of the visualization design and the willingness participants expressed to use them, we believe providing such visual analyses in online music services would be beneficial for both service users and providers.

\section{Conclusions}\label{sec_conc}
In this paper, we take a user-centric approach to the design of visual analyses of users' music history. A pilot study is conducted to investigate user needs in such visual analyses and collect users' opinions on some sketch design. We find that participants are interested in seeing their music listening patterns, and recommendations based on the selection of partial history are welcomed. We then propose four visualization plots: Bean plot, Transitional Pie plot, Instrument plot, and Calendar plot, and perform a user study to verify the utility of the three plots used for visual analyses. Results of the user study show the value of the proposed visualization in helping non-expert people doing visual analyses.

Our design and user study also identify some questions worthy of further study. How to make the relevances among music tracks understandable and useful to common users seems a challenge. How to evaluate the user experience of recommendations in our proposed Calendar plot is also our future work.

\bibliography{mybibfile}

\end{document}